\documentclass[preprint]{aastex61}
\usepackage{epstopdf}
\usepackage[toc]{appendix}
\usepackage{color}
\usepackage{lineno}
\usepackage{bm}

\graphicspath{{./}{figures/}}

\received{\today}
\submitjournal{ApJ}

\shorttitle{}
\shortauthors{Lukin et al.}

\begin{document}

\title{Triggering the magnetopause reconnection by solar wind discontinuities} 

\correspondingauthor{Alexander S. Lukin}
\email{as.lukin.phys@gmail.com}

\author{Alexander Lukin}
\affiliation{Space Research Institute, RAS, Moscow, Russia}
\affiliation{Faculty of Physics, National Research University Higher School of Economics, Moscow, Russia}

\author{Zhifang Guo}
\affiliation{School of Geophysics and Information Technology, China University of Geosciences, Beijing, China}
\affiliation{Physics Department, Auburn University, Auburn, AL, USA}

\author{Yu Lin}
\affiliation{Physics Department, Auburn University, Auburn, AL, USA}

\author{Evgeny Panov}
\affiliation{Space Research Institute, Austrian Academy of Sciences, Schmiedlstrasse 6, 8042 Graz, Austria}

\author{Anton Artemyev}
\affiliation{Department of Earth, Planetary, and Space Sciences and Institute of Geophysics and Planetary Physics, \\University of California, Los Angeles, CA, USA}
\affiliation{Space Research Institute, RAS, Moscow, Russia}

\author{Xiaojia Zhang}
\affiliation{Department of Physics, University of Texas at Dallas, Richardson, TX, USA}
\affiliation{Department of Earth, Planetary, and Space Sciences and Institute of Geophysics and Planetary Physics, \\University of California, Los Angeles, CA, USA}

\author{Anatoli Petrukovich}
\affiliation{Space Research Institute, RAS, Moscow, Russia}




\begin{abstract}
Magnetic reconnection is one of the most universal processes in space plasma that is responsible for charged particle acceleration, mixing and heating of plasma populations. In this paper we consider a triggering process of reconnection that is driven by interaction of two discontinuities: solar wind rotational discontinuity and tangential discontinuity at the Earth's magnetospheric boundary, magnetopause. Combining the multispacecraft measurements and global hybrid simulations, we show that solar wind discontinuities may drive the magnetopause reconnection and cause the mixing of the solar wind and magnetosphere plasmas around the magnetopause, well downstream of the solar wind flow. Since large-amplitude discontinuities are frequently observed in the solar wind and predicted for various stellar winds, our results of reconnection driven by the  discontinuity-discontinuity interaction may have a broad application beyond the magnetosphere.
\end{abstract}

\section{Introduction}

Magnetohydrodynamic (MHD) discontinuities are basic solutions for boundaries of plasmas with different properties \citep{bookLL:electrodynamics,Hudson70}, and are fundamental element of space plasma flows, e.g., solar wind \citep{Tsurutani&Ho99} and stellar winds \citep{Arons12,Sironi&Spitkovsky11}. Such discontinuities are considered as elementary blocks of MHD turbulence \citep{Servidio11:npg,Matthaeus15,Loureiro&Boldyrev17} and serve as a primary framework for magnetic field line reconnection, a major process responsible for the transfer of magnetic field energy to plasma heating and charged particle acceleration \citep{bookBirn&Priest07,book:Gonzalez&Parker,Pontin&Priest22}. Besides the spontaneous magnetic reconnection due to discontinuity instability \citep{Syrovatskii81,Loureiro07,Pucci14}, magnetic reconnection can be driven by an interaction of two discontinuities. The most investigated type of such driven reconnections is that due to interaction of a tangential discontinuity and a shock wave \citep[see examples of simulations for astrophysical and solar plasma systems in, e.g.,][]{Phan07:magnetosheath_reconnection,Sironi&Spitkovsky12, Matsumoto15, Omidi09, Guo18:magnetosheath_reconnection, Guo21:par_bowshock_reconnection,Guo21:perp_bowshock_reconnection}. Numerical simulations and theoretical models of discontinuity interactions associated with magnetic reconnection are largely motivated by multiple spacecraft observations in the near-Earth space environment and in the solar wind, the natural laboratory of plasma dynamics.  

Earth's bow shock and magnetopause are formed by solar wind interaction with the Earth’s dipole field and are separated by the magnetosheath, the region filled by subsonic (shocked) plasma flow. The magnetosheath-magnetosphere boundary, magnetopause, can be described as a tangential discontinuity \citep{deKeyser05}, an equilibrium current sheet where the geomagnetic field lines are 'closed'. Solar wind flow transport kinetic-scale discontinuities \cite[see statistical properties of such discontinuities in][]{Soding01,Vasquez07,Artemyev18:apj,Vasko21:apjl} sharing properties of both rotational and tangential discontinuities \citep{Neugebauer06,Artemyev19:jgr:solarwind} and carrying strong current density \citep{Podesta17,Greco16,Newman20}. Interaction of solar wind discontinuities with the bow shock significantly modifies the discontinuity structure \citep{Yan&Lee95,Lin96:discontinuity,Cable&Lin98} and can also produce various types of magnetosheath perturbations:  plasma bubbles \citep{Liu15:foreshock,An20:apj,Vu22}, density cavities \citep{Lin03,Lin&Wang05,Sibeck02,Sibeck21,Omidi13}, hot flow anomalies \citep{Lin97:hfa,Lin02:hfa,Omidi&Sibeck07,Omidi16}, and plasma pressure pulses \citep{Hubert&Harvey00, Farrugia18}. Both modification of the initial discontinuity configuration and generation of magnetosheath perturbations involve ion kinetic processes, and thus are well reproduced by hybrid simulations treating ions as particles and electrons as a fluid \citep[e.g.,][]{Lin96:discontinuity, Lin96:pressure_pulse,Lin&Wang05,Omidi16,Omidi20,Ng21:magnetopause_reconnection_by_jets,ChenLJ21:magnetopause_reconnection_by_foreshock_turbulence}. 

\begin{figure}
\centering
\includegraphics[width=0.5\textwidth]{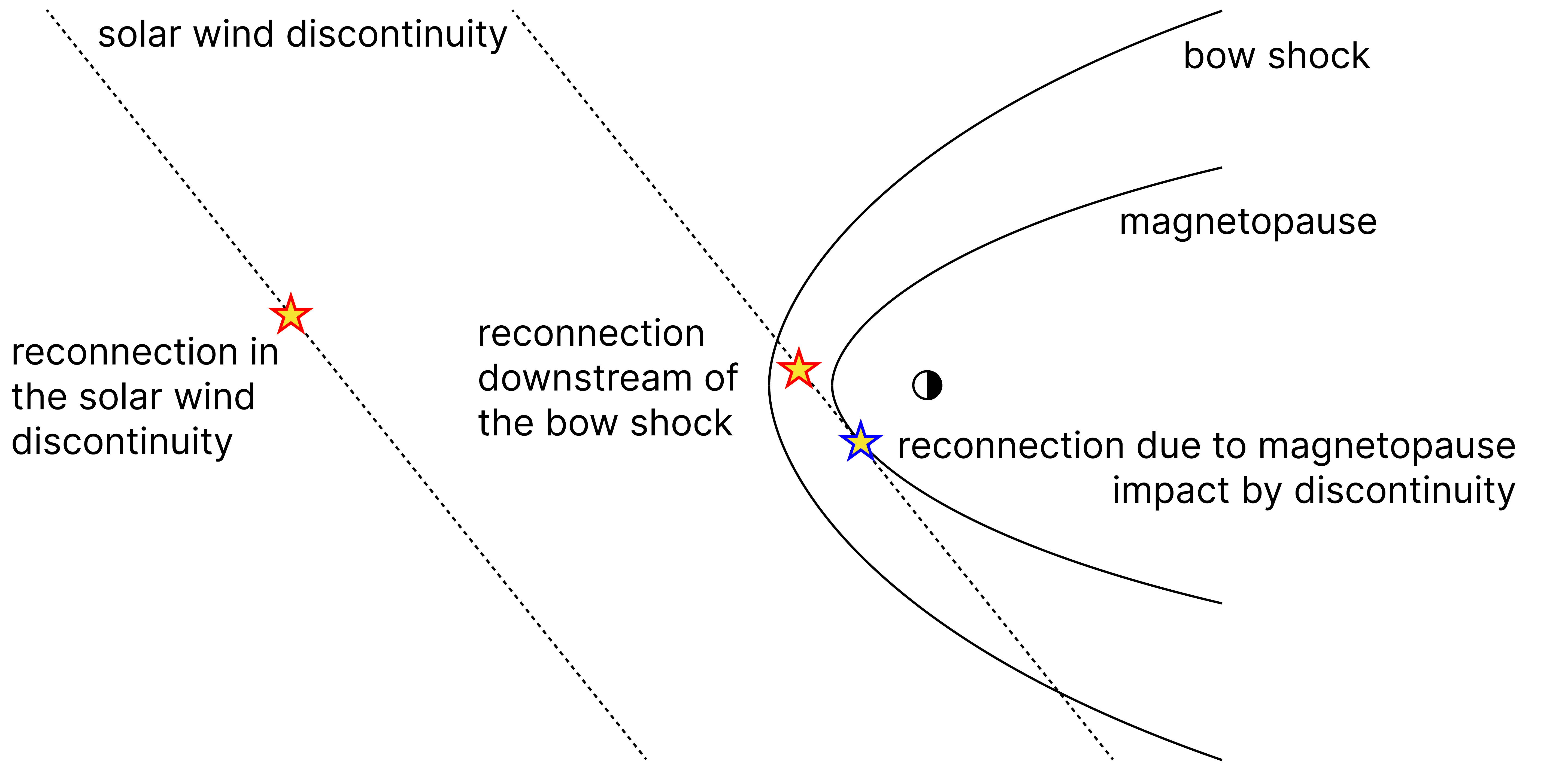}
\caption{Schematic view of main types of magnetic reconnection associated with the solar wind discontinuities. \label{fig1} }
\end{figure}

Schematic \ref{fig1} shows main types of magnetic reconnections observed within the day-side Earth's magnetosphere environment: reconnection at solar wind discontinuities \citep{Farrugia01,Gosling12,Phan06:reconnection}, reconnection due to the discontinuity interaction with the bow shock \citep{Pang10,Voeroes17,Wang19:reconnection_at_bow_shock,Gingell19,Gingell20,Guo21:perp_bowshock_reconnection}, and reconnection on the magnetopause \citep{Paschmann79,Paschmann13,Phan14,Burch16:science}. The latter one is generally driven by a change of the interplanetary magnetic field direction, i.e. by solar wind discontinuity arriving to the magnetopause. In this study we focus on this last type of reconnection associated with magnetopause interaction with magnetosheath perturbations, in which the sheath perturbations are generated by the solar wind discontinuity interaction with the bow shock. We consider very coherent perturbations, and study the magnetopause (that is tangential discontinuity, see \citet{Roth96,deKeyser05}) interaction with the solar wind rotational discontinuity modified by the passage through the bow shock (that is the shock-type discontinuity).

One of the important consequences of discontinuity interaction with the bow shock is its compression with the significant increase of the current density \citep{Phan07:magnetosheath_reconnection,Webster21,Kropotina21:ApJ}. Therefore, such amplified discontinuities should be more unstable to magnetic reconnection in the magnetosheath. Indeed, recent observations  \citep{Voeroes17,Wang19:reconnection_at_bow_shock,Gingell19,Gingell20} and kinetic simulations \citep{Guo18:magnetosheath_reconnection} show reconnection signatures downstream of the Earth’s bow shock, and such reconnection is associated with  the tangential discontinuity interaction with the bow shock. However, reconnection does not necessary destroy the current layer, which can reach the magnetopause. Rotational discontinuities are especially stable to magnetic reconnection (see Discussion for more details). Interaction between the modified rotational discontinuity and the magnetopause, that is a tangential discontinuity, may also drive the magnetic field line reconnection and magnetosheath-magnetosphere plasma exchange. This type of reconnection is in focus of our study. 

Roles of various magnetosheath perturbations in the magnetopause dynamics is well documented by in-situ spacecraft measurements \citep[e.g.,][]{Sibeck00:magnetopause_motion,Jacobsen09:magnetopause_motion&HFA,Archer14:magnetosheath_flows&magnetopause_motion,Hietala18:jet} and ground-based magnetometers \citep[e.g.,][]{Archer15}. Moreover, such dynamics results in the generation of a broad range of ultra-low-frequency (MHD) waves within the Earth’s magnetosphere, and these waves are usually detected by the ground-based and spacecraft magnetic field and plasma measurements \citep[e.g.,][]{Wang18:ULF&foreshock_transients,Wang20:ULF&foreshock_transients,Shi21:foreshock_processes}. Such magnetosheath perturbations may propagate from the day-side to midtail magnetosheath \citep{Wang18:ARTEMIS&transient_foreshock_perturbations} and drive the flank magnetopause dynamics \citep{Wang20:ARTEMIS&foreshock_impact_magnetopause}. Therefore, we shall consider possibility of both day-side and night-side (flank) magnetopause reconnection driven by the solar wind discontinuity crossing the bow shock. 

In this study we combine the 3D global hybrid simulation \citep[see model description in][]{Lin&Wang05} of the interaction between the magnetopause and the solar wind discontinuities with THEMIS \citep{Angelopoulos08:ssr}, MMS \citep{Burch16}, and ARTEMIS \citep{Angelopoulos11:ARTEMIS} observations to investigate the subsequent discontinuity interaction with the magnetopause and driven magnetopause reconnection. Our main event describes the day-side reconnection, and for this event we use the simulation results to confirm our interpretation of ARTEMIS and THEMIS measurements. The second event describes ARTEMIS and MMS observations of night-side magnetopause reconnection driven by a solar wind discontinuity. We use this event to demonstrate that discontinuity-magnetopause interaction may drive magnetic reconnection in different plasma flow configurations. We also analyze three supplementary events with magnetopause crossings before and after the solar wind discontinuity impact, which confirm that the solar wind discontinuity impact may trigger magnetic reconnection at the initially non-reconnected magnetopause.

\section{Spacecraft observations}

\subsection{Data and methods}
Below we show two events of  interaction between solar wind rotational discontinuities and the Earth's magnetopause. We use the data obtained onboard THEMIS/ARTEMIS \citep{Angelopoulos08:ssr,Angelopoulos11:ARTEMIS} and MMS \citep{Burch16} missions. THEMIS/ARTEMIS fluxgate magnetometers \citep{Auster08:THEMIS} and electrostatic analyzers (ESA; \citet{McFadden08:THEMIS}) provide us with the magnetic field vector, moments of ion and electron distribution functions (number density, temperatures and bulk velocities) and phase space density measurements with a spin time resolution (3 and 4 second for THEMIS and ARTEMIS respectively). MMS fluxgate magnetometer \citep{Russell16:mms} measurements are available with 1/16s resolution, while the plasma moments measured by Fast Plasma Investigation (FPI) instrument \citep{Pollock16:mms} are provided with $4.5$ seconds resolution.

To restore the local (lmn) coordinate system of the discontinuity we use the minimum variance analysis (MVA;  \citet{bookISSI:Sonnerup}). To estimate the spatial scale of the discontinuity we use almost the same single-spacecraft technique as proposed by \citet{Haaland&Gjerloev13}, but since the MVA often does not distinguish between the intermediate and normal directions we use the velocity component perpendicular to the main magnetic field ($l$) component, $\boldsymbol{v_{mn}}=\boldsymbol{v}-(\boldsymbol{v}\cdot \boldsymbol{e_l})\boldsymbol{e_l}$ in our analysis, i.e. we may overestimate the discontinuity thickness $L=\Delta t |\boldsymbol{v_{mn}}|$ ($\Delta t$ is the time interval of discontinuity crossing by spacecraft).


For steady reconnection process under symmetrical boundary conditions and a finite guide field, there exist two symmetrical rotational discontinuities bounding the outflow region, and thus we expect that the variation of the plasma flow velocity across the outflow region discontinuities should satisfy the Walen relation $\Delta\boldsymbol{v}=\pm \Delta \boldsymbol{B}/(4\pi\rho)$ \citep{Parker57,Petschek64,Lin&Lee93:magnetopause}. The magnetopause reconnection however, are usually characterized by asymmetrical boundary conditions \citep{Phan07:magnetosheath_reconnection,Phan14} so the jump of plasma flow speed should be comparable to the asymmetric Alfven speed $v_{Al,asym}$ \citep{Cassak&Shay07,Swisdak&Drake07}:
\begin{equation}
    v_{Al,asym}^2=\frac{B_{l1}B_{l2}(B_{l1}+B_{l2})}{4\pi(\rho_{1}B_{l2}+\rho_{2}B_{l1})},
\end{equation}
where $B_{l1,2}$, $\rho_{1,2}$ are the values of anti-parallel reconnecting magnetic field component and plasma density on both sides of the outflow region boundaries. In our selected events, spacecraft do not cross the reconnecting magnetopause, but observe the discontinuities likely connecting to the magnetopause reconnection side. Therefore, we may not estimate the magnetic field configuration of the reconnecting magnetopause, and we use component-wise estimates of the asymmetrical Alfven speed $v_{Aj,asym}^2=(B_{j1}B_{j2}(B_{j1}+B_{j2}))/(4\pi(\rho_{1}B_{j2}+\rho_{2}B_{j1}))$, where $j=x,y,z$ (we use GSM coordinate system through the paper), and $B_{j1}$, $\rho_1$ and $B_{j2}$, $\rho_2$ are measured in the magnetosheath near the discontinuity and in the magnetosphere.

\subsection{Day-side event}
\begin{figure}
\centering
\includegraphics[width=0.75\textwidth]{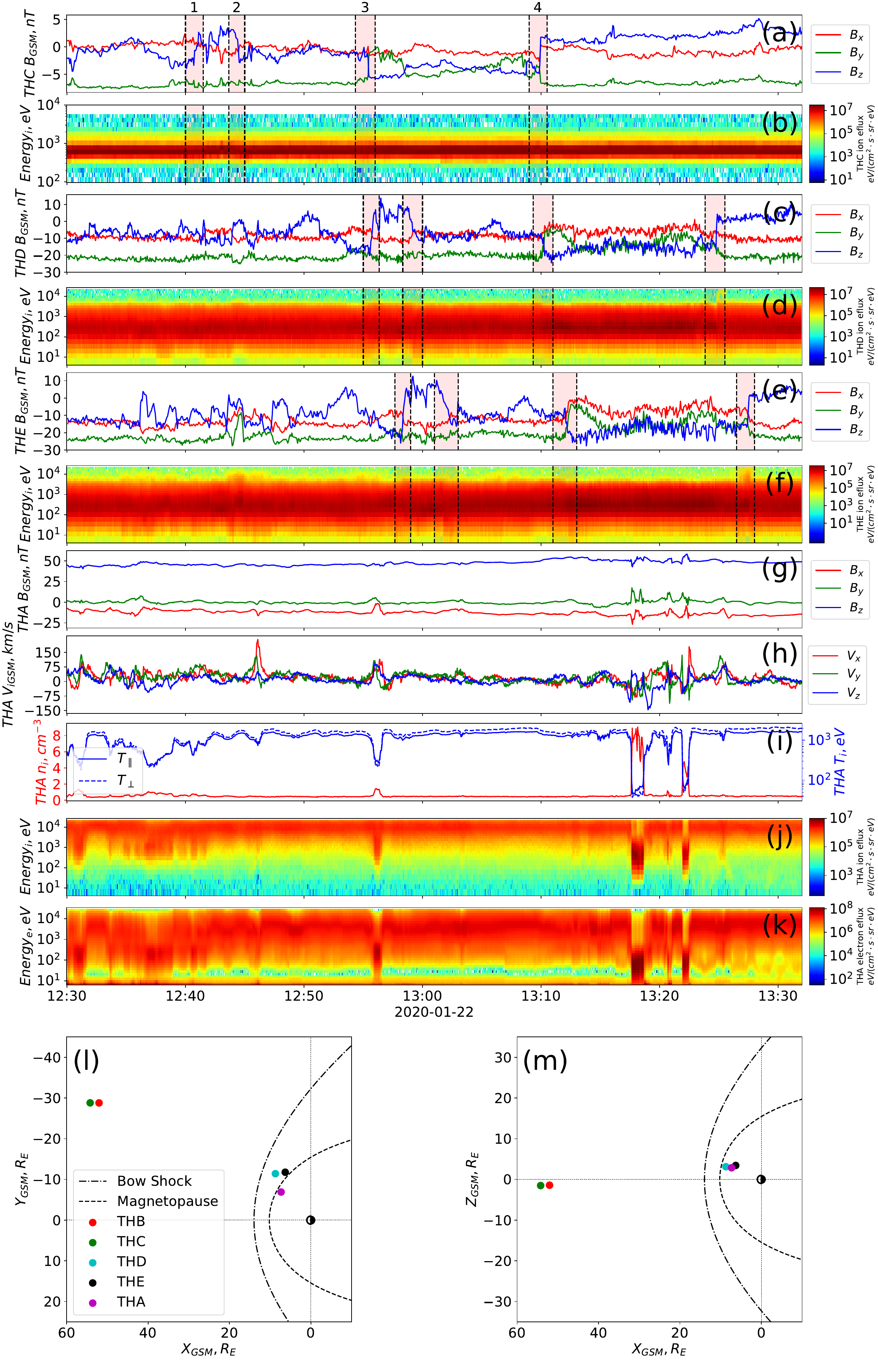}
\caption{The overview of THEMIS observations on 2020-01-22. (a),(b): solar wind discontinuity (magnetic field and ion spectrum on THEMIS C), (c)-(f): the same discontinuity observed downstream of the bow shock (magnetic fields and ion spectra on THEMIS D\&E), (g)-(k) magnetospheric observations (magnetic fields, plasma flow velocities, electron density and temperature, ion and electron spectra on THEMIS A). Panels (l) and (m) show spacecraft location in $XY$ and $XZ$ GSM planes relative to the model bow shock \citep{Wu00:bowshock} and magnetopause \citep{Shue97:magnetopause}. \label{fig2}}
\end{figure}

Figure \ref{fig2} shows an overview of the event with the day-side magnetopause reconnection driven by the solar wind discontinuity. THEMIS C (ARTEMIS P2) is in the solar wind, upstream of the bow shock (see panels ($l$), ($m$)), and observes a series of the solar wind discontinuities (shaded in red in Fig. \ref{fig2}(a),(b)) at $\sim$12:41, $\sim$12:44, $\sim$12:55 and $\sim$13:10 with the magnetic field rotation mostly seen in  GSM $B_x$, $B_z$ for the first and second events, and in GSM $B_y$, $B_z$ for  in the third and fourth events. The magnetic field rotation is accompanied by jumps of solar wind flow $\Delta v_{x,y,z}\approx B_{x,y,z}/\sqrt{4\pi\rho}$ (not shown) with the almost constant field magnitude $|B|$. These are typical properties of rotational discontinuity \citep{Tsurutani&Ho99,Neugebauer06}. Although single spacecraft measurements do not allow to check the normal magnetic field component with necessary accuracy \citep[see][]{Knetter04}, we use the time delay of discontinuity measurements by two ARTEMIS spacecraft to improve accuracy of MVA \citep[see details of this method in][]{Artemyev19:jgr:solarwind} and estimate a normal magnetic field component: $B_n/\Delta B_l\approx (0.2, 0.1, 0.1, 0.05)$ for four discontinuities shown in Fig. \ref{fig2}. For three first discontinuities $B_n$ is quite comparable to the $B_l$ change across the discontinuity (the discontinuity amplitude), and thus we can confirm that these are rotational discontinuities. Their spatial scales (thicknesses) in the solar wind, based on THEMIS C observations, are $L_1\sim 7500$km, $L_2\sim6200$km, $L_3\sim3050$km and $L_4\sim2200$km.


These rotational discontinuities cross the bow shock and are observed downstream (in the magnetosheath) by THEMIS D and E. The interaction with the bow shock compresses the discontinuity and increases the magnetic field amplitude from $\sim 5$nT in the solar wind to $\sim 20-30$nT in the magnetosheath. The spatial scales of these discontinuities in the magnetosheath, based on THEMIS D and E observations, are $L_{1d}\sim 2000$km, $L_{2d}\sim5300$km, $L_{3d}\sim1500$km,  $L_{4d}\sim550$km and $L_{1e}\sim 2500$km, $L_{2e}\sim7600$km, $L_{3e}\sim2000$km,  $L_{4e}\sim800$km. Such intensification of discontinuities downstraeam of the bow shock has been reproduced in simulations \citep[e.g.,][]{Yan&Lee95,Lin96:discontinuity,Guo18:magnetosheath_reconnection,Guo21:perp_bowshock_reconnection,Guo21:par_bowshock_reconnection,Kropotina21:ApJ} and reported in spacecraft observations \citep[e.g.,][]{Webster21,Kropotina21:ApJ}. The interesting aspect of the discontinuity observation in the magnetosheath is the population of energetic ions (mostly seen around the last discontinuity). Energy spectrum clearly shows this population at $>5$keV range, significantly higher than the shocked solar wind energies. The pitch-angle distribution of this population indicates that energetic ions are moving along magnetic field lines (not shown), i.e. along $-y$ direction from the magnetopause. Therefore, this may be a leakage of the magnetopsheric population (for which $5-15$keV energy range is expected). Such observations in the magnetosheath suggest that discontinuity are connected to the magnetopause reconnection region. The scenario for such observations includes the discontinuity impact on the magnetopause, local reconnection between the magnetosheath and magnetospheric field lines, and ion leakage along the reconnected magnetic field lines (along the discontinuity surface) to the magnetosheath, where this magnetospheric ion population is observed in association with the discontinuities. An alternative, or rather supplementary mechanism can be the magnetospheric plasma leakage into the magnetosheath due to magnetosphere compression by the pressure pulse formed by the discontinuity interaction with the bow shock \citep[see][]{Lin96:pressure_pulse,Archer12}. Such a compression may release magnetospheric plasma into magnetosheath and may drive the magnetopause reconnection \citep[e.g.,][]{Hietala18:jet}.

Before the discontinuity arrives to the magnetopause, THEMIS A was inside the magnetosphere (the magnetopause crossing from the magnetosheath to the magnetosphere occurs at $\sim$12:05) and observed almost stationary magnetic field, stagnant plasma, low plasma density, and hot ions (before 13:17). Then THEMIS A detected several injections of the cold dense magnetosheath plasma into the magnetosphere: increase of density, low-energy population in ion and electron spectra, magnetic field perturbations. These injections of cold plasma are accompanied by observations of plasma jets seen as bipolar flows around $\sim$12:57, $\sim$13:18 and $\sim$13:22. The component-wise estimates of asymmetric Alfven speed calculated for all marked discontinuties are $v_{Ax,asym}\sim 50-70$km/s, $v_{Ay,asym}\sim 20-30$km/s, $v_{Az,asym}\sim 90-120$km/s and comparable to the observed jet speeds. 
Therefore, we may suggest that discontinuity's impact on the magnetopause triggers magnetic reconnection that is responsible for the magnetosheath plasma penetration into the magnetosphere and the magnetospheric plasma penetration into the magnetosheath. This impact (and magnetic reconnection) also drives MHD waves propagating away from the magnetopause and observed by THEMIS A as quasi-periodic $v_x$ variations modulating the cold ion and electron populations after 13:22:00 \citep[see discussion of such waves in][]{Hartinger13,Zhang20:jgr:ulf,Wright&Elsden20}.

\begin{figure}
\centering
\includegraphics[width=1\textwidth]{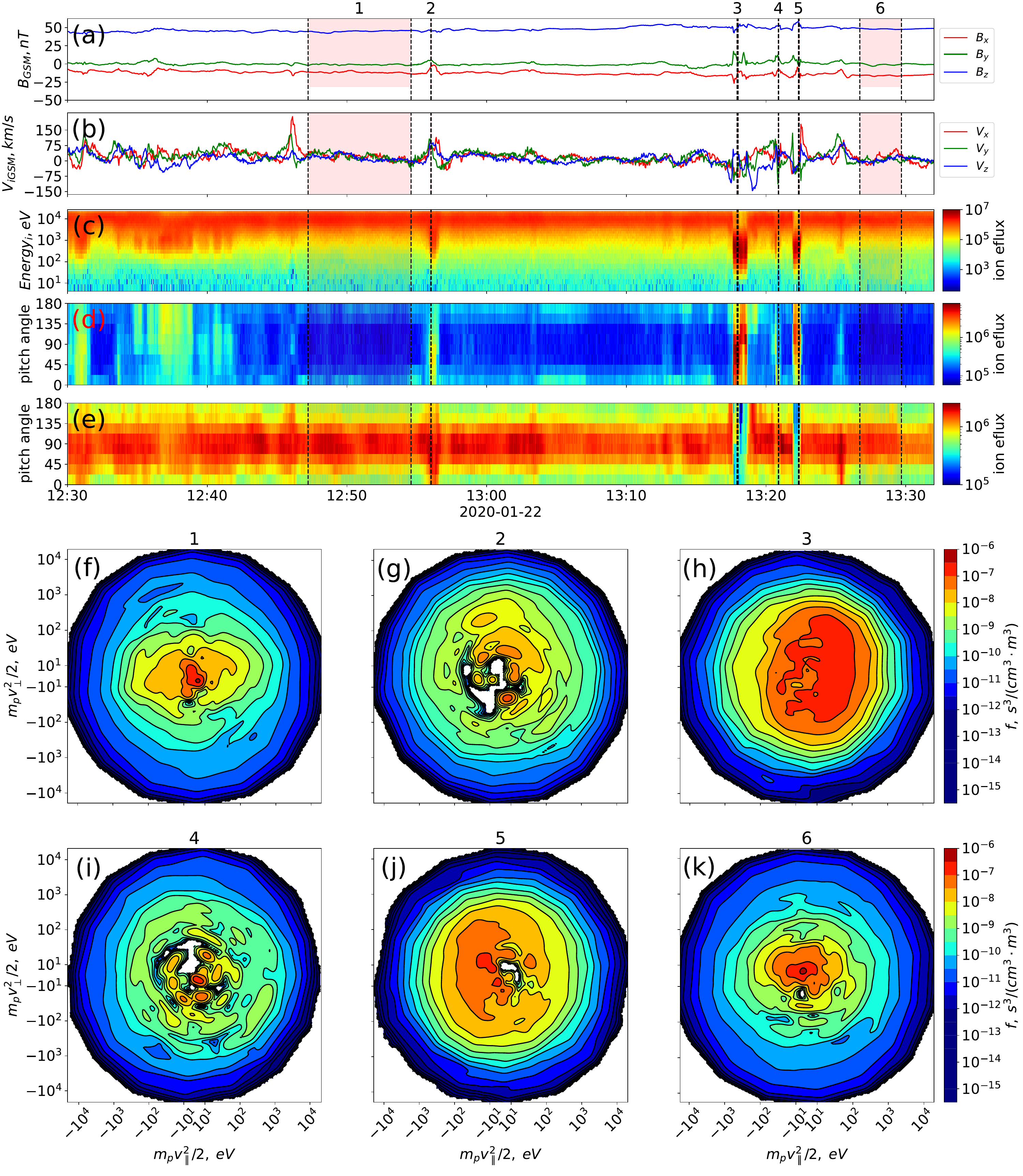}
\caption{The overview of the solar wind discontinuity interaction with the magnetopause observed by THEMIS A: magnetic field components (a), ion flow components (b), ion energy spectrum (c), ion pitch-angle distributions for different energy ranges ($[100-2000]$ eV for (d), and $[4-20]$ keV for (e)), 2D cuts of ion distributions in ($m_pv_\parallel^2/2$, $m_pv_\perp^2/2$) plane for several intervals indicated by numbers.  \label{fig3} }
\end{figure}

To investigate details of the discontinuity interaction with the magnetopause, we zoom in THEMIS A observations and add information about ion kinetics in Fig. \ref{fig3}. Pitch-angle distributions of thermal ($<2$keV) and hot ($\in[4,20]$ keV) ions show that the bipolar $v_z$ signature around 13:17-13:19 consists of cold (magnetosheath) ions flowing parallel and hot (magnetosheric) ions flowing anti-parallel to the background magnetic field. This result resembles well the bipolar (reversal) plasma flows from the asymmetric magnetoause reconnection \citep[see review by][and references therein]{Paschmann13}, i.e. THEMIS A may have crossed the outflow region of reconnection. Comparison of velocity distributions collected before 12:55 (subinterval \#1), during observations of $v_z>0$ (subintervals \#2 and \#3), and well after the bipolar $v_z$ (subinterval \#6) shows that magnetospeheric plasma (subinterval \#1,6) are predominantly field-aligned anisotropic, whereas injected magnetosheath plasma (subintervals \#2, \#3 and \#5) is transversely anisotropic (both these types of anisotropy are quite typical for the magnetopshere and magnetosheath, see \cite{Gary95,Samsonov07:angeo} showing that magnetosheath plasma generally has $T_\perp/T_\parallel>1$, and \cite{Wang13:ions} showing the magnetospheric thermal plasma often has $T_\parallel>T_\perp$). Therefore, Fig. \ref{fig3} further confirms that THEMIS A observed signatures of magnetic reconnection at the magnetopause: bipolar $v_z$ pulses and injection of cold magnetotheath plasma into the magnetosphere. This reconnection triggers the MHD (ultra-low-frequency) waves that quasi-periodically brings the magnetosheath population (likely entered into magnetosphere through the reconnected magnetopause) to THEMIS A location via ${\bf E}\times{\bf B}$ drift, as shown in the cold transversely anisotropic population in the distribution functions measured during subinterval \#2-5. 

\subsection{Night-side event}
\begin{figure}
\centering
\includegraphics[width=0.75\textwidth]{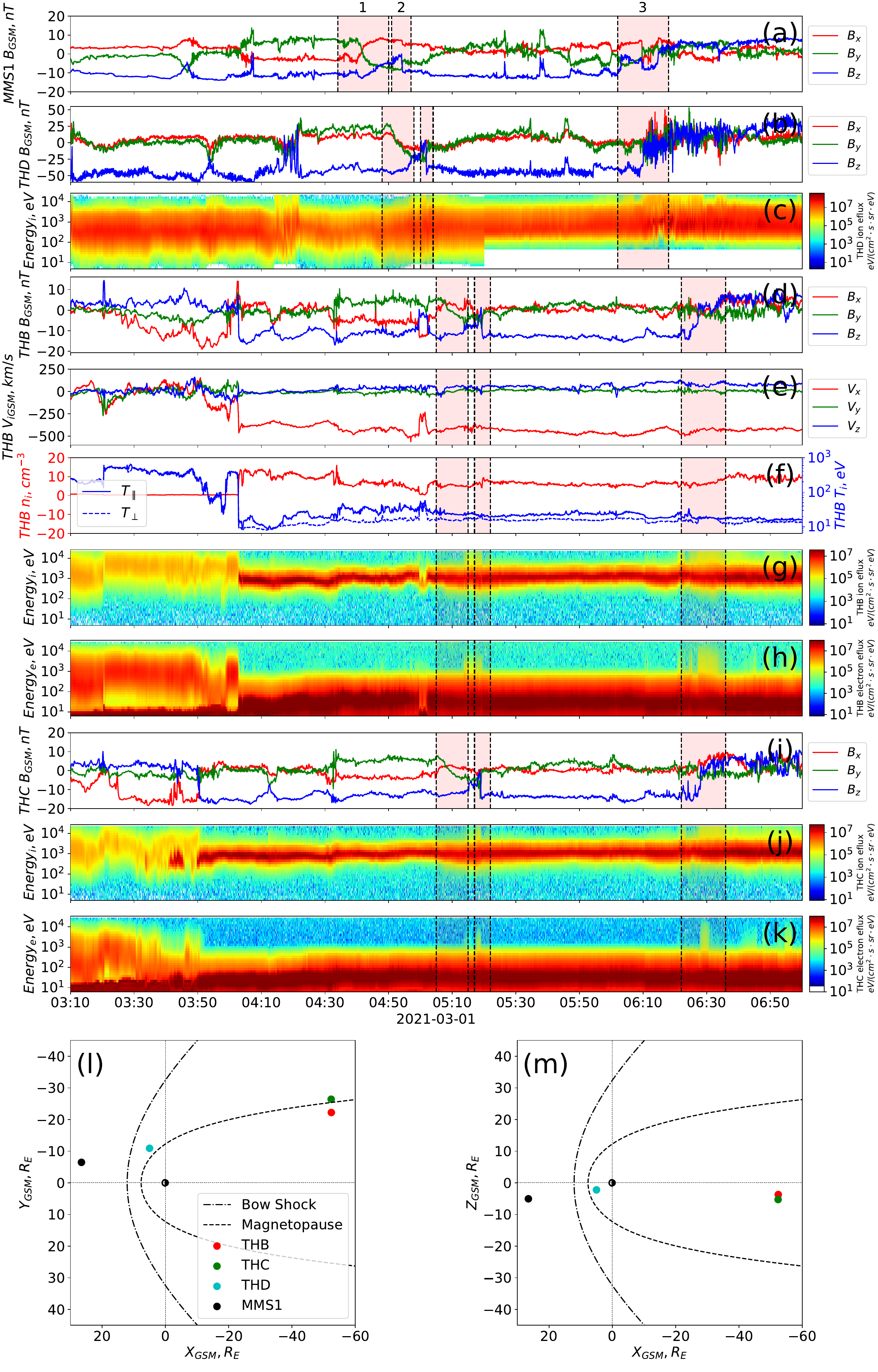}
\caption{The overview of MMS\&THEMIS\&ARTEMIS observations on 2021-03-01. From top to bottom: (a) solar wind discontinuity (magnetic field on MMS\#1), the same discontinuity observed downstream of the bow shock (magnetic fields (b) and ion spectra (c) on THEMIS D), magnetic fields observed downstream of the flank bow shock  ((d) for THEMIS B \& (i) for THEMIS C), plasma flow velocities on THEMIS B (e), electron density and temperature on THEMIS B (f), ion and electron spectra on ((g,j) for THEMIS B \& (h,k) for THEMIS C). Bottom panel shows spacecraft location relative to the model bow shock and magnetopause.  \label{fig4} }
\end{figure}

To investigate a possibility for a solar wind discontinuity to trigger the night-side (flank) magnetopause reconnection, we consider MMS and THEMIS observations during 2021-03-01 event shown in Fig. \ref{fig4}. MMS upstream of the bow shock observes several solar wind discontinuities: large scale magnetic field rotation between 04:35 and 04:50 (marked as 1 in panel (a) of Fig. \ref{fig4}), smaller scale rotation between 04:52 and 04:56 (marked as 2) and a series of $B_z$ reversals between 06:00 and 06:20 (marked as 3). The spatial scale of the first discontinuity estimated from MMS1 data is about $10^5$ km. In fact this structure consists of several embedded discontinuities with spatial scales of 11700, 23500 and 9800 km. The second discontinuity have a spatial scale of about 7400 km and consists of two adjacent discontinuities with estimated scales of 2000 and 1900 km. The last marked interval includes several $B_z$ reversals, the first one, observed at 06:03, have a spatial scale of about 31000 km and consists of two smaller scale discontinuities with scales of 6200 and 1800 km. The one, observed at 06:15, has spatial scale of 5800 km and doesn't appear in the magnetosheath and nigh-side observations. 

Each discontinuity is accompanied by solar wind velocity jump (not shown). The same series of discontinuities is observed by THEMIS D downstream of the bow shock in the dayside magnetosheath. Discontinuities are compressed at the bow shock: the magnetic field increases from $10$nT in the solar wind to $\sim 40$nT in the dayside magnetosheath (THEMIS D). Spatial scale decreases to 35000 km ($2500 + 4000 + 6800$ km) for the first group of discontinuities, 1200 km for the second discontinuity, and 18000 km ($4200 + 4800$ km) for the third group of discontinuities. Note the downstream of the day side bow shock is further compressed by plasma flow breaking at the magnetoapuse, and this results in large discontinuity magnitude increase \citep[see discussion in][]{Kropotina21:ApJ}. Thus, it would be interesting to check the discontinuity effect on the night side magnetosheath \citep[see examples confirming such propagation of solar wind discontinuities to the night-side magnetopause in, e.g.,][]{Wang18:transients_midtail,Wang20:transients_midtail}, where magnetopause surface is almost parallel to the plasma flow \cite{Hasegawa12,Lukin19}. 

THEMIS B and C (ARTEMIS P1 and P2) show several discontinuities in the night side magnetosheath (downstream from the bow shock, $\sim 50$ Earth radii away from the day side THEMIS D observations). These discontinuities are only weakly compressed by the bow shock: magnetic field magnitude increases from $10$nT to $15$nT and discontinuity spatial scales stay almost the same or even increase:  $10^5$ km ($11000 + 17000 + 12400$ km) for the first group of discontinuities, 9000 km for the second discontinuity, and 64000 km ($22000 + 15000$ km) for the third group of discontinuities. Note a typical ion thermal gyroradius in the shocked solar wind (in the magnetosheath) is $\sim100$km for cold ions and $\sim 1000$km for hot (magnetospheric) ions. Thus small-scale discontinuities are ion kinetic structures for hot ions, but is quite thick for cold magnetosheath ions.

\begin{figure}
\centering
\includegraphics[width=0.9\textwidth]{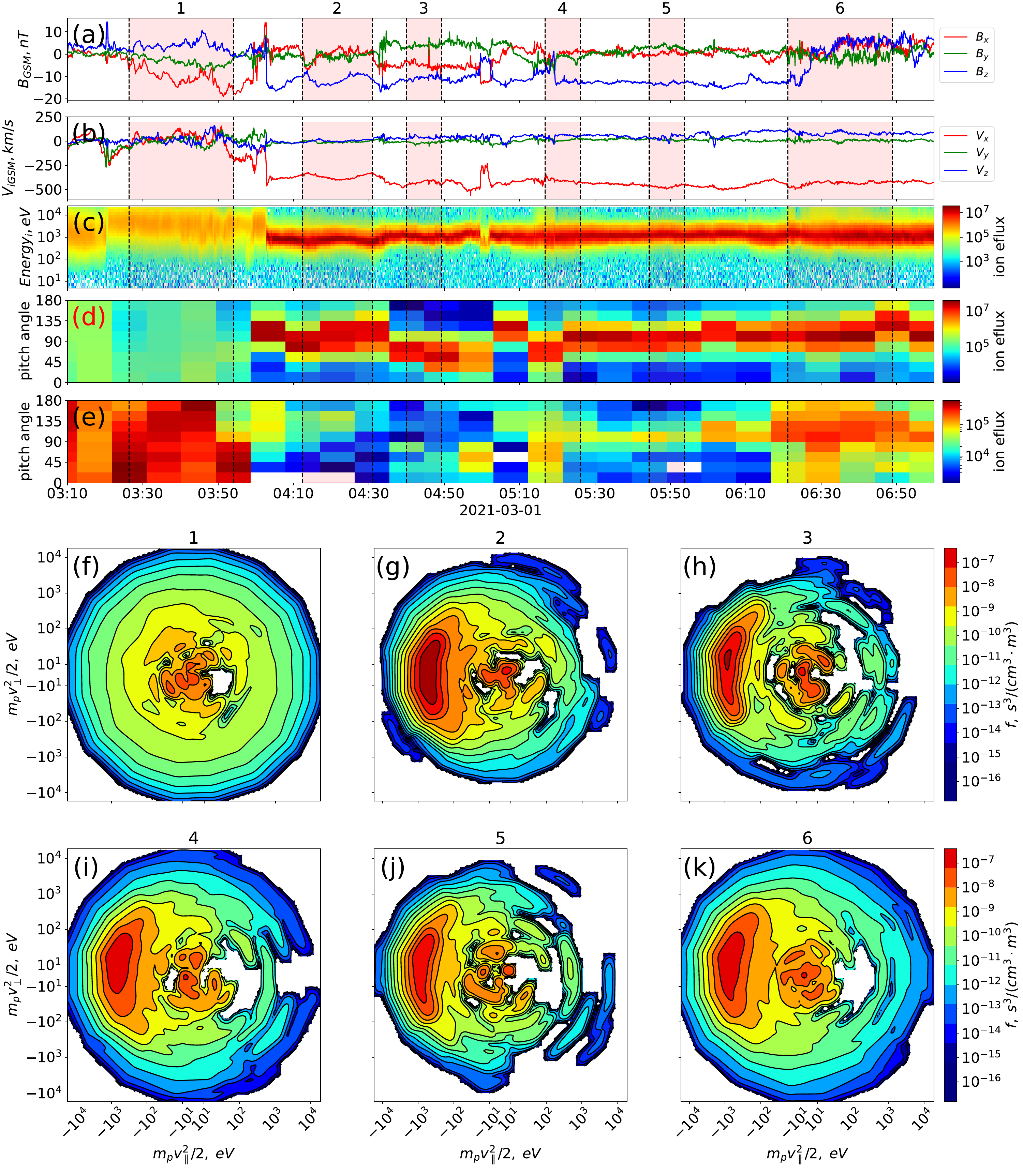}
\caption{The overview of the solar wind discontinuity interaction with the magnetopause (THEMIS B): magnetic field components (a), ion flow components (b), ion energy spectrum (c), ion pitch-angle distributions for different energy ranges ($[100-2000]$ eV for (d) and $[4-20]$ keV for (e)), 2D cuts of ion distributions in ($m_pv_\parallel^2/2$, $m_pv_\perp^2/2$) plane for several intervals indicated by numbers.  \label{fig5}}
\end{figure}

THEMIS C and B cross the magnetopause at 03:50 and 04:05 (a clear change in ion spectra is seen in Fig. \ref{fig4}(g,j)), and after the magnetopause crossing both THEMIS B and C stably measure magnetosheath plasma. However, there are several bursts of  hot magnetospeheric ion population on the magnetosheath side around discontinuities, between 05:12 and 05:20, 06:20 and 07:00. In Fig. \ref{fig5} we describe details of these hot plasma bursts associated with discontinuities. For reference we use ion velocity distributions measured within the magnetosphere (interval \#1 in Fig. \ref{fig5}(f)) and in the quiet magnetosheath (intervals \#2,3,5 in Fig. \ref{fig5}(g,h,j)): the magnetospheric population is characterized by a finite $\sim 10$ keV fluxes isotropically distributed in $(\parallel,\perp)$ energy space, whereas the magnetosheath population is characterized by large amplitude fluxes of field-aligned flow ($E_\parallel \sim 1$keV). During interval \#4 (the first group of discontinuities) ion distribution function shows a combination of magnetosheath (flowing) and magnetospheric (hot isotropic) plasmas. Pitch-angle distributions in Fig. \ref{fig5}(d,e) confirms that discontinuities are associated with appearance of hot isotropic ion population. A similar picture is observed for the third group of discontinuities (interval \#6): pitch-angle distributions and velocity distribution show an appearance of a population of hot ($\sim 10$ keV) isotropic magnetosphereic plasma (see Fig. \ref{fig5}(d,e, k)). Observation of magnetosphereic plasma far from the magnetopause (deep inside the magnetosheath) in association with discontinuities suggest that THEMIS C\&B cross field lines connecting to the reconnected magnetopause, i.e. there is a magnetic reconnection in regions where these discontinuities impact to the magnetopause.

\subsection{Supporting events}

Events in Figs. \ref{fig2}, \ref{fig4} demonstrate the magnetopause reconnection after the solar wind discontinuity impact, but for these events we do not have observations of the non-reconnected magnetopause before the  arrival of the discontinuity. Thus, the reconnection might have already started even before the discontinuity arrives, and would be only modified by this impact. To verify the key role of the discontinuity impact in triggering the reconnection, we analyze three supporting events when the spacecraft crossed the magnetopause before and after the discontinuity arrives. In all three events the magnetopause does not show reconnection signatures before the discontinuity impact, but becomes reconnected only afterwards.

\begin{figure}
\centering
\includegraphics[width=1\textwidth]{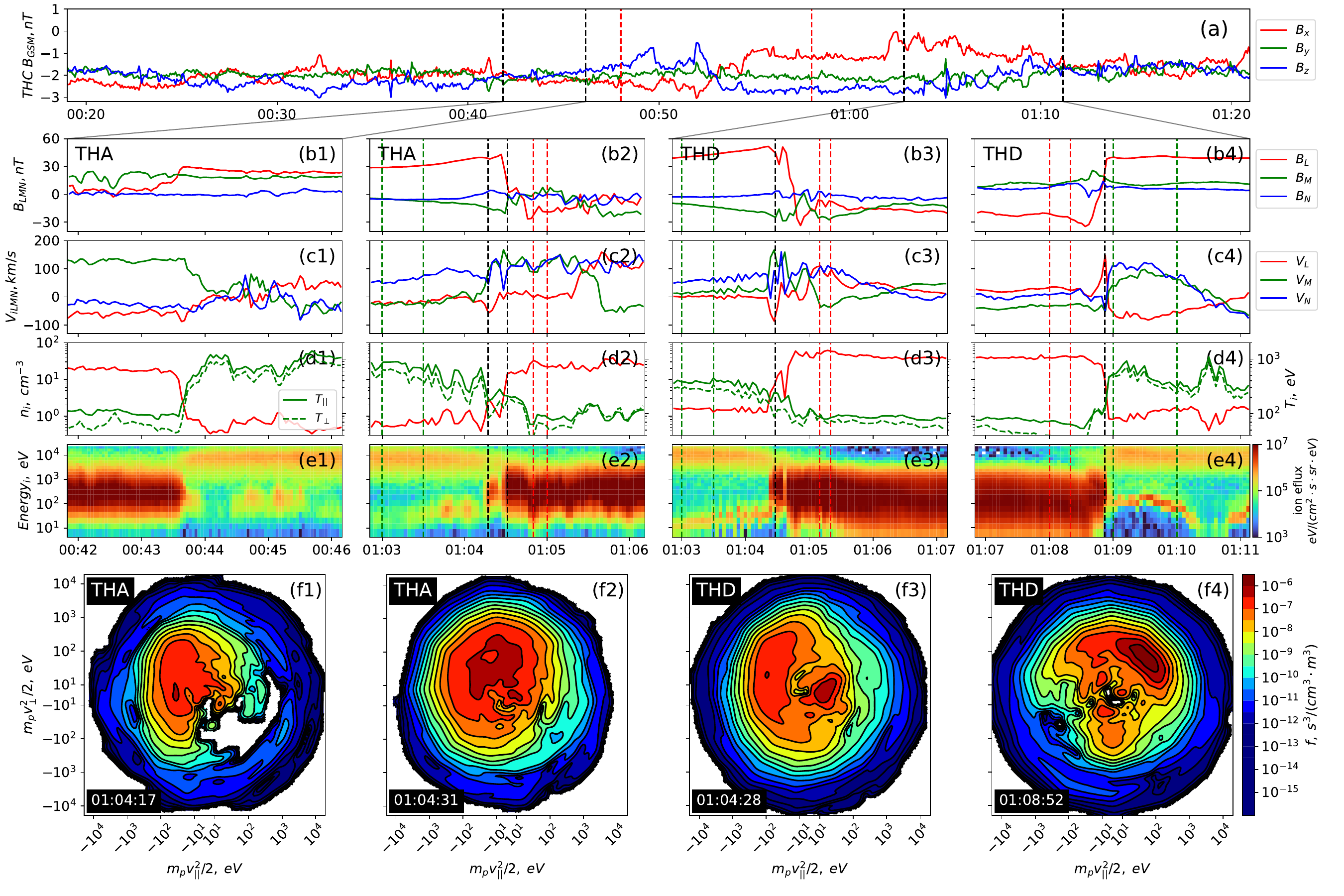}
\caption{Overview of THEMIS\&ARTEMIS observations on 2019-09-20. From top to bottom: (a) solar wind discontinuity shown by two red dashed lines (magnetic field at ARTEMIS\#2); magnetic field of the magnetopause crossings by THEMIS A and D before (b1) and after (b2-b4) the solar wind discontinuity impact; plasma flow velocities (c), plasma density and ion temperature (d), ion spectra (e) for the magnetopause crossings before (1) and after (2-4) the solar wind discontinuity impact; the corresponding ion velocity distributions around the magnetopause (f1-f4). \label{fig4add}  }
\end{figure} 

Figure \ref{fig4add} shows an overview of the 2019-09-20 event. THEMIS C is at $r_{GSM}\sim (-24, -57,  12) R_E$ and observes a series of discontinuities at $\sim$00:48-00:55 UT. During 00:20-01:20 UT, THEMIS A crosses the magnetopause several times, and two such crossings are shown in panels (b1-e1) and (b2-e2). As shown in the figure, during the first crossing before the discontinuity arrives at the magnetopause, there are no ion jets or other signatures of the reconnection, i.e., THEMIS A crosses non-reconnected magnetopause.

In the second crossing right after the discontinuity arrives at the magnetopause, THEMIS A is at $r_{GSM}\sim (9,  9,  2) R_E$ and observes the bifurcated current sheet with several ion jets, the primary signatures of the magnetic reconnection. 
In the local coordinate system, the observed ion bulk velocity change across the magnetopause at two times, as marked by the black dashed lines in panels (b2-e2), are $\Delta v_{1obs}\sim (-56, -34, -79)$ km/s and $\Delta v_{2obs}\sim (5,  25, -65)$ km/s. Magnitudes of these velocities, however, are much smaller than magnitudes of velocity changes estimated from the model of the reconnected magnetopause \cite{Phan13,Hudson70,Paschmann86}:
\[
    \Delta\boldsymbol{v_{pred}}=\boldsymbol{v_2}-\boldsymbol{v_1}=\pm\sqrt{\frac{1-\alpha_1}{4\pi\rho_i}}\left(\boldsymbol{B_2}(1-\alpha_2)/(1-\alpha_1)-\boldsymbol{B_1}\right)
\]
where $\alpha=\frac{4\pi(p_{\parallel}-p_{\perp})}{B^2}$ is the anisotropy factor, $p_{\parallel}, \; p_{\perp}$ are the plasma pressures parallel and perpendicular to $\boldsymbol{B}$. Subscripts 1 and 2 denote the magnetosheath inflow region (marked by red dashed lines in Fig. \ref{fig4add}(b-e)) and the jet region, respectively. For this crossing, $\Delta v_{1pred}\sim (-250, 81, -16)$ km/s and $\Delta v_{2pred}\sim (230, -107, 3)$ km/s. Moreover, the observed velocity changes are also smaller than the estimate of the asymmetric Alfven velocity, $V_{Al,asym}\sim 133.05$ km/s. Therefore, despite that quasi-parallel ion flows are observed in the ion velocity distribution (see panels (f1-f2)), the observed and predicted values of ion velocity changes are much different, and THEMIS A may not observe the reconnected magnetopause. 

After the discontinuity arrives at the magnetopause, both THEMIS D and THEMIS E (not shown) observe almost the same signatures of magnetic reconnection, though much more evident at THEMIS D. Both spacecraft are in the magnetosphere near the magnetopause at $r_{dGSM}\sim (9, 7, 1) R_E$ and $r_{eGSM}\sim (5,  12,  1) R_E$, and due to the magnetosphere compression by the discontinuity both spacecraft appear in the magnetosheath after the discontinuity (panels (b3-e3)). The estimated magnetopause thickness for the two crossing shown in panels (b3-e3) and (b4-e4) are $L_{d1} \sim 924$ km ($\sim 2.38 \rho_{iMSH}$, where $\rho_{iMSH}=\frac{m_i c v_T}{eB}$ is the averaged value of the ion thermal gyroradius in the magnetosheath) and $L_{d1} \sim 821$ km ($\sim 2.26 \rho_{iMSH}$) \citep[see detailed description of the method in][]{Haaland14}. Therefore, THEMIS D indeed crosses a ion-scale magnetopause current sheet. In the local coordinate system, the observed and predicted ion velocity changes across the magnetopause (at the two times marked by the black dashed lines in panels (b3-e3) and (b4-e4)) are $\Delta v_{1obs}\sim (-175,  200,   -9)$ km/s, $\Delta v_{1pred}\sim (-173,   -1,    6)$ km/s and $\Delta v_{2obs}\sim (115, -21, -59)$ km/s, $\Delta v_{2pred}\sim (167,   7,   -12)$ km/s. The estimated asymmetric Alfven velocity is $V_{Al,asym}\sim 124.88$ km/s. Moreover, quasi-parallel flows are seen in the ion velocity distribution (panel (f3)) during the first THEMIS D crossing. During the second THEMIS D crossing, the strong ion jet is tangential to the magnetopause, but at the moment of the jet observation the main magnetic field component is around zero, i.e., there is a large uncertainty in projecting to the local magnetic field direction. During both crossings, the nonzero normal velocity components are observed together with the low energy population (panels (e3-e4)). Therefore, THEMIS D observations clearly demonstrate the magnetopause reconnection after the discontinuity impact, whereas THEMIS A observations confirm that the magnetopause was not reconnected before this impact.

\begin{figure}
\centering
\includegraphics[width=1\textwidth]{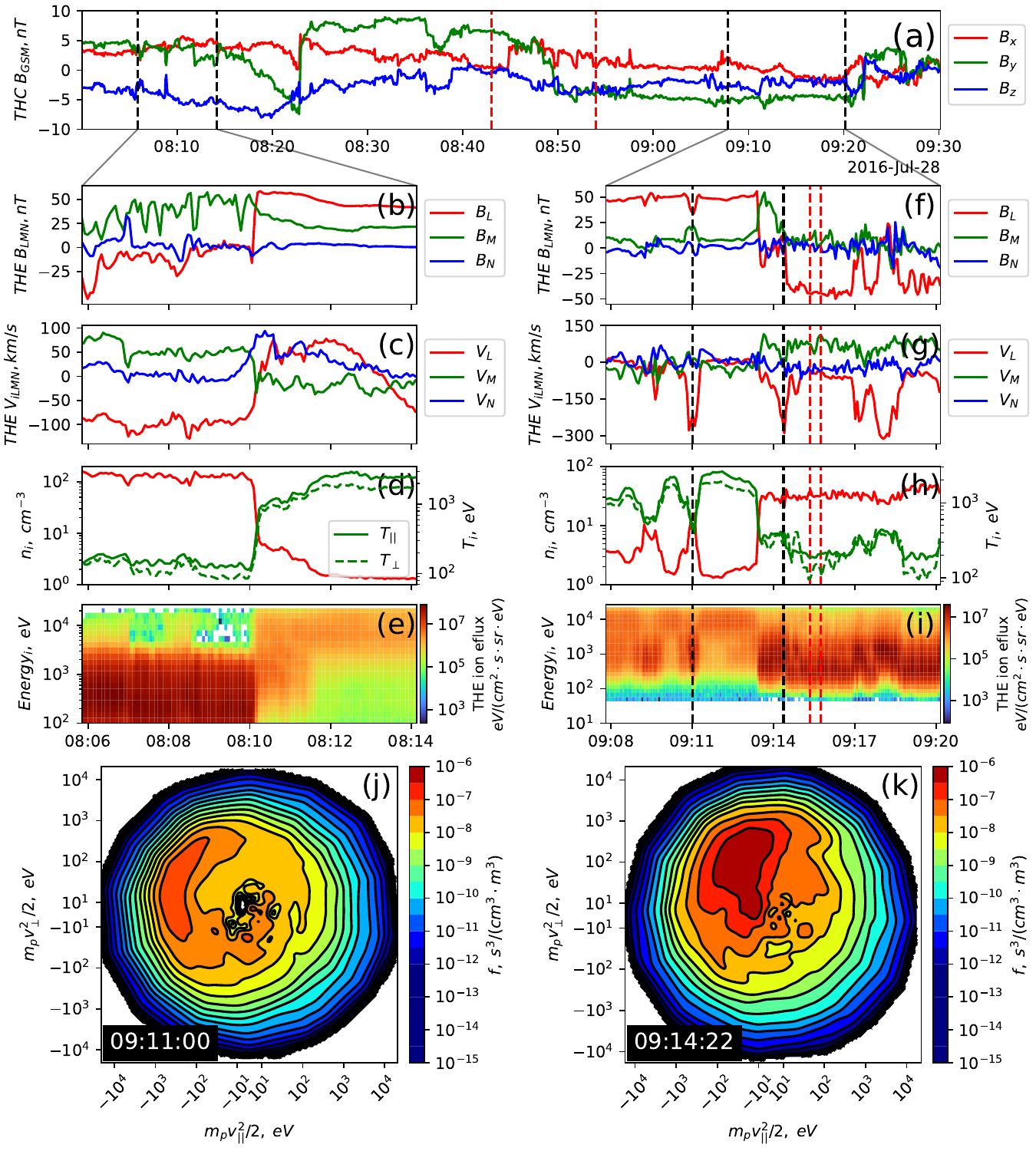}
\caption{Overview of THEMIS\&ARTEMIS observations on 2016-07-28. From top to bottom: (a) solar wind discontinuity as marked by the two red dashed lines (magnetic field at ARTEMIS\#2); magnetic field of the magnetopause crossings by THEMIS E before (b) and after (f) the solar wind discontinuity impact; plasma flow velocities (c and g), plasma density and ion temperature (d and h), ion spectra (e and i) for the magnetopause crossings from (b and f); ion velocity distributions around the magnetopause crossing after the solar wind discontinuity impact (j,k). \label{fig2add} }
\end{figure} 

Figure \ref{fig2add} shows another example of the solar wind discontinuity interaction with the magnetopause on 2016-07-28. During this event, THEMIS C is in the solar wind at $r_{GSM}\sim (20, -52, 16) R_E$ and observes a series of discontinuities at $\sim$ 08:43-08:54 UT. The same series of discontinuities are observed by THEMIS D in the magnetosheath around 09:00-09:05 UT (not shown). Within 08:06-09:30UT, THEMIS A crosses the magnetopause twice. The first crossing is shown in panels (b-e), and for this crossing the observed and predicted flow changes are $\Delta v_{obs}\sim (163, -58,  91)$ km/s and $\Delta v_{pred}\sim (100, 1047, -14)$ km/s, whereas the estimated asymmetric Alfven velocity is $V_{Al,asym}\sim 16$ km/s. The large difference in the observed and predicted ion velocity change indicates that THEMIS A crosses the non-reconnected magnetopause.

Panels (f-i) show the second crossing, after the discontinuity arrives to the magnetopause. Starting from 09:00UT (when the discontinuity is detected in the magnetosheath by THEMIS D), THEMIS A observes variations in the magnetopsheric magnetic field, whereas during the magnetopause crossing THEMIS A detects the bifurcated current sheet and several ion jets (the primary signatures of the magnetic reconnection). The observed and estimated ion velocity changes for the two times marked by black dashed lines in panels (f-i) are $\Delta v_{1obs}\sim (-206,  -63,   22)$ km/s, $\Delta v_{1pred}\sim (-332,  -31,  -11)$ km/s and $\Delta v_{2obs}\sim (-251,  -53,   10)$ km/s, $\Delta v_{2pred}\sim (-211,  1,  2)$ km/s. The estimated asymmetric Alfven velocity is $V_{Al,asym}\sim 236.73$ km/s. Therefore, the similar velocity change between observations and the model prediction suggests that THEMIS A indeed crosses the reconnected magnetopause. Moreover, panels (j-k) show the corresponding 2D slices of the ion distribution functions with clear quasi-parallel ion jets. Thus, these observations show that the interaction between the solar wind discontinuity and the magnetopause can trigger magnetic reconnection of the initially non-reconnected magnetopause.

\begin{figure}
\centering
\includegraphics[width=1\textwidth]{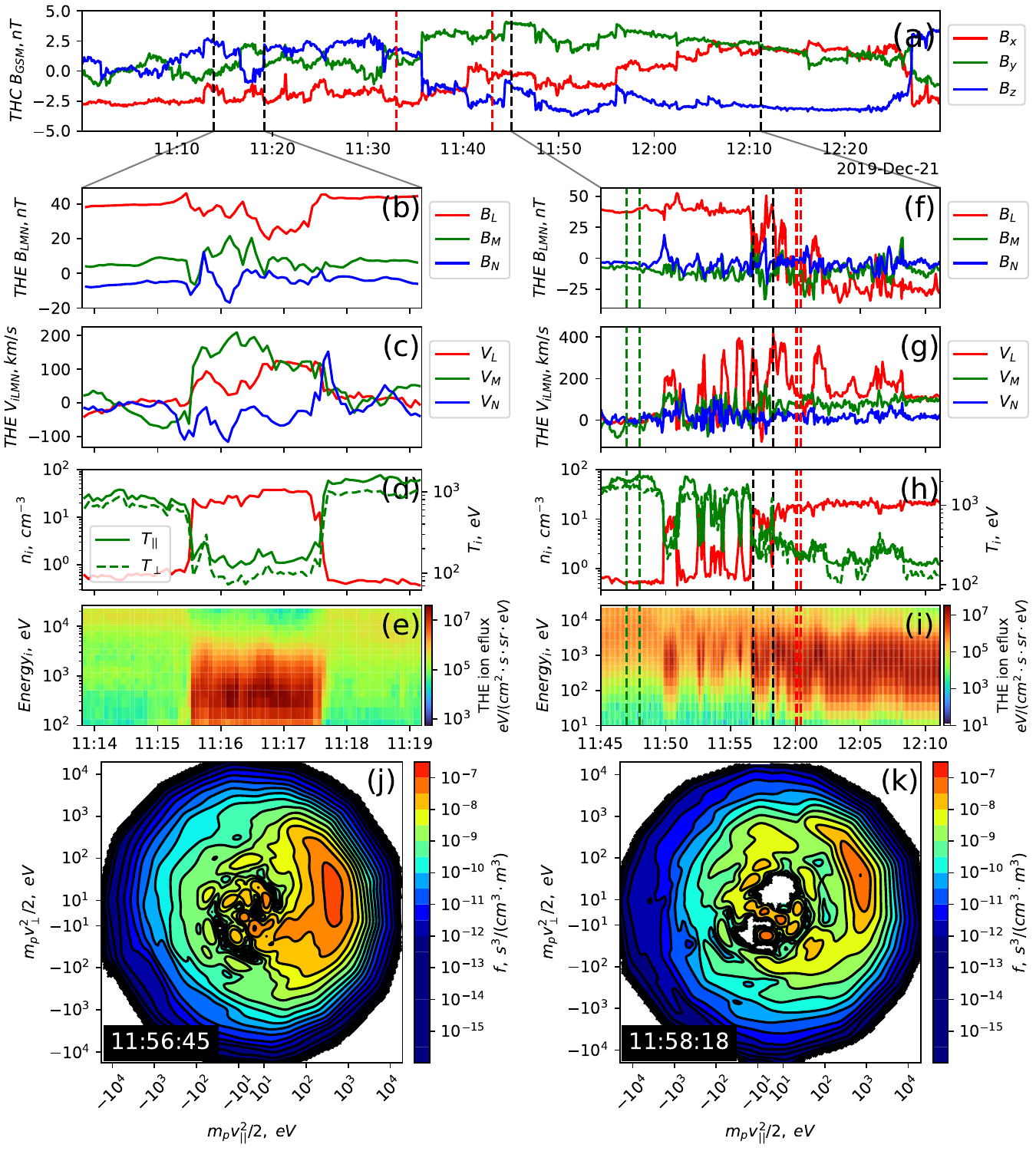}
\caption{Overview of THEMIS\&ARTEMIS observations on 2019-12-21. From top to bottom: (a) solar wind discontinuity shown by two red dashed lines (magnetic field at ARTEMIS\#2); magnetic field of the magnetopause crossings by THEMIS E before (b) and after (f) the solar wind discontinuity impact; plasma flow velocities (c and g), plasma density and ion temperature (d and h), ion spectra (e and i) for the magnetopause crossings from (b and f); ion velocity distributions around the magnetopause crossing after the solar wind discontinuity impact (j,k). \label{fig15add} }
\end{figure}

The third supplementary event is shown in Figure \ref{fig15add}. During this event, THEMIS C is in the solar wind at $r_{GSM}\sim (28, -51,  15) R_E$ and observes a series of discontinuities during $\sim$11:33-11:43 UT. The same series of discontinuities are observed by THEMIS A in the magnetosheath around $\sim$11:30 UT (not shown).  Within 11:10-12:20 UT, THEMIS E crosses the magnetopause several times. The first and the second crossings are shown in panels (b-e). During these crossings, no ion jets or other reconnection signatures are detected. Panels (f-i) show the third crossing, after the discontinuity arrives at the magnetopause. During the magnetopause crossing, THEMIS E is at $r_{GSM}\sim (11,  -4, 5) R_E$ and observes variations in the magnetic field accompanied by ion jets. The observed and estimated ion velocity changes for the two times marked by black dashed lines in panels (f-i) are $\Delta v_{1obs}\sim (204, -82, -41)$ km/s, $\Delta v_{1pred}\sim (225,  61, -20)$ km/s and $\Delta v_{2obs}\sim (238, -29,  -2)$ km/s, $\Delta v_{2pred}\sim (251,  16, -25)$ km/s. The estimated asymmetric Alfven velocity is $V_{Al,asym}\sim 173$ km/s. Panels (j-k) show the corresponding 2D slices of the ion distribution functions with quasi-parallel ion jets. As in events from Figs. \ref{fig4add}, \ref{fig2add}, the observed ion velocity changes are comparable to the predicted changes and to the value of the asymmetric Alfven velocity, whereas ion velocity distributions show clear field-aligned ion jets. Therefore, Fig. \ref{fig15add} confirms results from Figs. \ref{fig4add}, \ref{fig2add} that the solar wind discontinuity impact may trigger the reconnection of the initially non-reconnected magnetopause.

\section{Global hybrid simulation}
In this section, a 3-D global hybrid code of \citet{Lin&Wang05,Tan11:YuLin,Guo18:magnetosheath_reconnection,Guo20:magnetopause_reconnection} is applied to investigate the evolution of the solar wind discontinuity and understand the dayside magnetopause reconnection driven by the impact of the discontinuity. In this hybrid model, ions are considered as fully kinetic  particles, while electrons are taken to be a massless fluid. A spherical coordinate system $(r,\theta,\phi)$ is employed. The simulation domain covers the dayside region ($x\geq 0$) with $r\in[3.5R_E,25R_E]$, where $R_E$ is the Earth radius. A perfectly conducting boundary condition is applied for the inner boundary at $r=3.5R_E$. The inflow boundary is located at $r=20R_E$, where the solar wind flows toward the Earth together with the interplanetary magnetic field (IMF). The initial IMF is assumed to be $(-1.25,-3.96,-3.86)$nT, the solar wind ion number density is set as $N_0=12.0{\rm cm}^{-3}$, and the ion temperature is chosen as $6.7$eV, based on the conditions observed by THEMIS C  before the rotational discontinuity arrives (the first, day-side event). The IMF on the sunward side of the discontinuity is chosen as $(-0.24,-5.45,1.52)$ nT. The solar wind velocity changes across the rotational discontinuity from $(-356.4,-27.2,-3.87)$ km⁄s to $(-350.0,17.7,-30.37)$ km⁄s. More detailed descriptions of how to impose discontinuity in the hybrid model can be found elsewhere \citep{Guo18:magnetosheath_reconnection, Guo21:par_bowshock_reconnection}. The number of grid cells is $n_r\times n_\theta\times n_\phi=320\times220\times180$, and a total of $\sim 5\times10^9$ particles are used. To reduce computing expenses, the grid cells in the radial direction are set to be non-uniform. A smaller grid size of $\Delta r\simeq0.025 \to 0.05R_E$ is used around the magnetopause and in the magnetosheath, while a larger grid size is set in the solar wind. We also set  the solar wind ion inertial length equal to $0.1$Re, i.e., a scaling factor is about $\sim 9.57$ is needed to scale the time and velocities to the realistic values.

\begin{figure}
\centering
\includegraphics[trim={0 4.5cm 0 0},clip,width=1\textwidth]{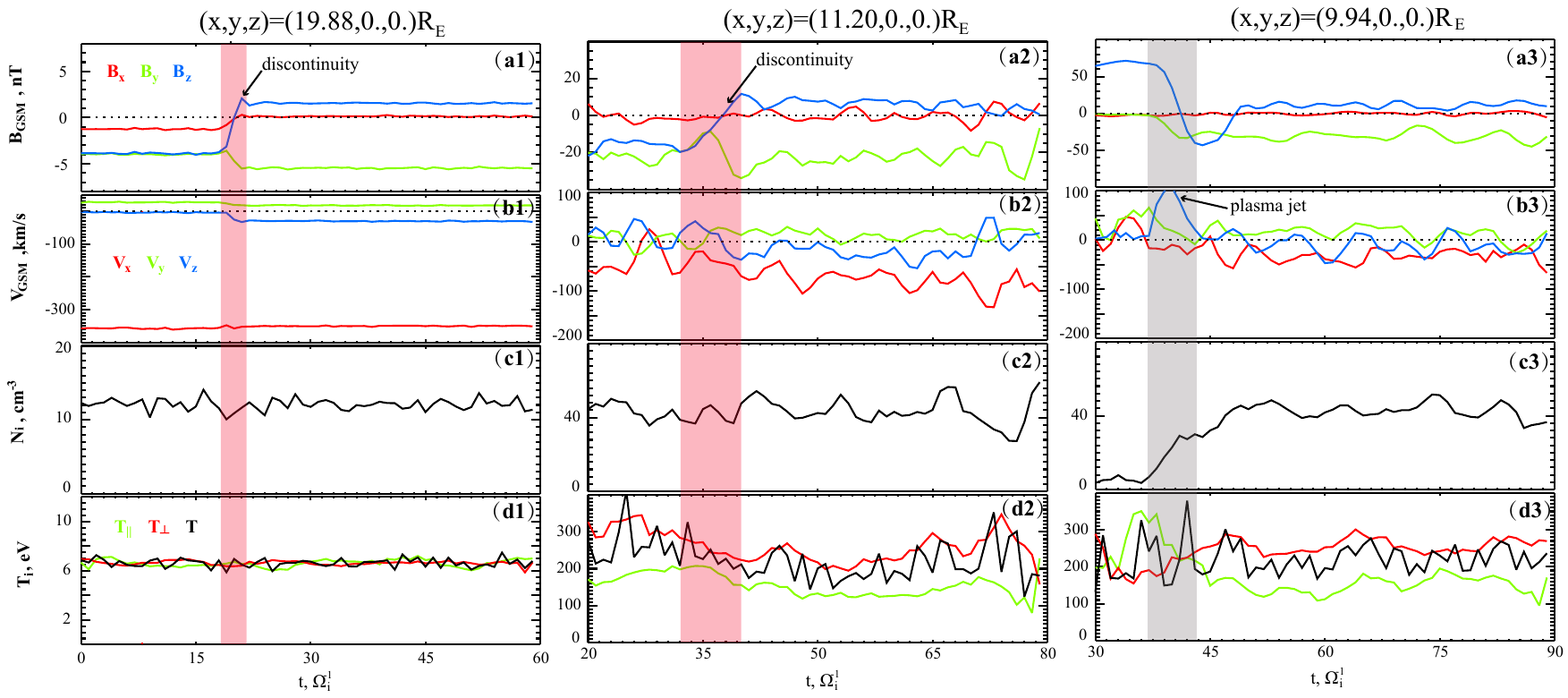}
\caption{The overview of global hybrid simulation of solar wind discontinuity (left panels), the same discontinuity in the downstream of the bow shock (center panels), and magnetopause reconnection (right panels): magnetic field vector (a1-a3),  ion velocity components (b1-b3), ion number density (c1-c3),  and ion temperature (d1-d3). \label{fig6} }
\end{figure}

Figure \ref{fig6} shows the discontinuity in the numerical simulation at three time intervals. The initial solar wind discontinuity (left panels) repeats the configuration and magnetic field components of the ARTEMIS observations from Fig. \ref{fig2}, a rotational discontinuity with a constant plasma density and jumps of plasma velocity across the discontinuity. The discontinuity downstream of the bow shock (center panels) is characterized by a magnetic field increase from the initial $\sim 5$nT to $\sim 20$nT. This magnetic field compression resembles the THEMIS A observations in the magnetosheath (see Fig. \ref{fig2}). In the magnetosheath, the ion number density $N_i$ (Fig. \ref{fig6}c2) and ion temperature $T_i$ (Fig. \ref{fig6}d2) increase, and the ion flow brakes with the ion bulk velocity $V_i$ (Fig. \ref{fig6}b2) slows down. When the discontinuity arrives at the magnetopause (right panels), the simulation shows typical signatures of magnetic reconnection: plasma jet in $V_z$ (Fig. \ref{fig6}b3) and an increase of ion temperature (Fig. \ref{fig6}d3) due to magnetospheric plasma leakage (compare with THEMIS A observations shown in Fig. \ref{fig3}).

\begin{figure}
\centering
\includegraphics[trim={0cm 0cm 0cm 0},clip,width=0.85\textwidth]{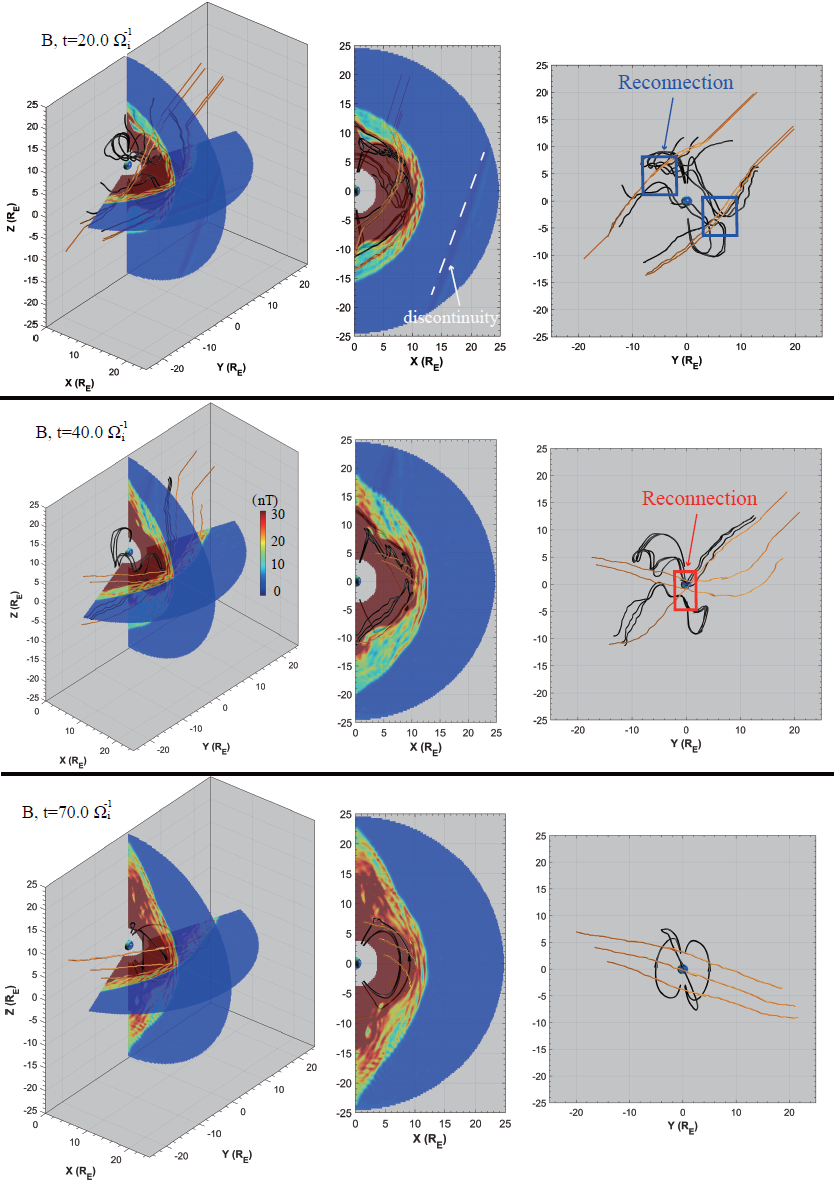}
\caption{The overview of solar wind discontinuity interaction with the magnetosphere: magnetic field strength in 3D and 2D planes. Three moments are shown: at $t\sim20\Omega_i^{-1}$ the RD was in the solar wind, at $t\sim 40\Omega_i^{-1}$ the RD interacts with the magnetopause, at $t\sim 70\Omega_i^{-1}$ the RD passes through the magnetopause. 
\label{fig7} }
\end{figure}

To further confirm that the discontinuity interaction with the magnetopause triggers magnetic reconnection, we visualize the solar wind discontinuity motion. The contour plots of Figure \ref{fig7} show the magnetic field strength for three moments of time: before arriving the solar wind discontinuity (top panels), when solar wind discontinuity interacts with the magnetopause (middle panels), and when solar wind discontinuity passes through the magnetopause (bottom panels). In the figure, the orange lines mark the magnetic field lines outside the magnetopause, and the black lines denote the field lines at the magnetopause. At $t=20$, before the solar wind discontinuity arrives at the magnetopause, magnetopause flux ropes has formed by reconnection between the initial IMF $(-1.25, -3.96, -3.86)$ nT and the dipole field, which are denoted by the black field lines. It is found that the magnetopause reconnection occurs in the high-latitude regions of the dayside magnetopause on the dawnside and duskside, which is marked by the blue boxes (note this region were not observed by spacecraft). When the solar wind discontinuity interacts with the magnetopause, at $t=30$, magnetic reconnection is seen at low- and middle latitudes of the dayside magnetopause, which is denoted as the red box. Finally, when the solar wind discontinuity passes through the magnetopause, no reconnection is found at the dayside magnetopause due to the magnetic field turns from $(-1.25, -3.96, -3.86)$ nT to $(-0.14,-5.45,1.52)$ nT.


\begin{figure}
\centering
\includegraphics[trim={5cm 3cm 5cm 3cm},clip,width=1\textwidth]{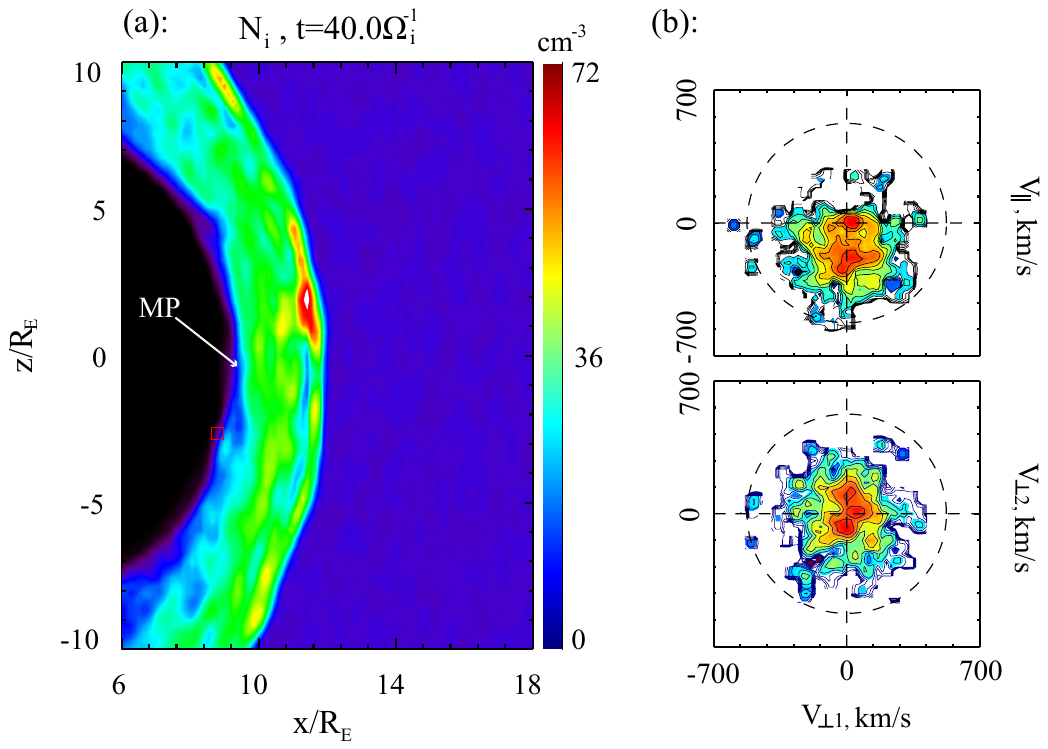}
\caption{(a) A zoom-in view of the ion number density in the y=0 plane; (b) Ion velocity distributions in the ($V_{\perp,1}$,$v_\parallel$) and ($V_{\perp,1}$,$V_{\perp,2}$) planes on the magnetosphere side at the location marked by the red box in Fig. 8a.  \label{fig8} }
\end{figure}

Using advantage of the hybrid simulation which treats ions as particles, we plot ion distribution functions right around the region of the magnetopause impact by the solar wind discontinuity. Figure \ref{fig8} shows that distributions on the magnetosphere side contain two ion populations: the hot field-aligned flowing population and cold population. The latter one penetrates from the magnetosheath along reconnected magnetic field lines. These model distributions resemble observational distributions \#3,5 from Fig. \ref{fig3}, which are measured when cold magnetosheath plasma injections are observed within the magnetosphere. Therefore, simulation results shown in Fig. \ref{fig6}-\ref{fig8} confirm the scenario of the magnetic reconenction driven by the magnetopause impact by solar wind discontinuity. 

\section{Discussion and Conclusions}
We presented observational and simulation evidence that the solar wind discontinuity impact may trigger the magnetic reconnection at the magnetopause. More general, the rotational discontinuity is intensified by crossing the shock wave (the Earth bow shock) and then impacts the tangential discontinuity (magnetopause), triggering the magnetic reconnection and plasma mixing (cross-penetration of magnetosheath and magnetospheric plasmas). We note that the rotational discontinuity interaction with the bow shock significantly modifies the discontinuity properties and add compressional perturbations, and thus we shall rather discuss interaction of the tangential discontinuity (magnetopause) with the discontinuity sharing properties of both rotational and tangential discontinuities. Reconnection of such rotational discontinuity downsteram of the bow shock is possible when the bow shock is quasi-parallel \citep{Guo21:par_bowshock_reconnection} or quasi-perpendicular, with large magnetic field rotation angle \citep{Guo21:perp_bowshock_reconnection}. In our cases, the bow shock configuration does not support the discontinuity reconnection in the magnetosheath, and thus discontinuity reaches the magnetopause and triggers the magnetopause reconnection. Although this scenario has been considered for the specific plasma system of the solar wind interaction with the Earth's magnetosphere, we may speculate about an application of the similar discontinuity-discontinuity interaction in other systems.

Discontinuity interactions should be rather common phenomena for solar wind (and more general, stellar wind) where such discontinuities may be formed by Alfven wave turbulence \citep[e.g.,][]{Buti98,Vasquez&Hollweg99,Vasquez&Hollweg01} and nonlinear Alfven wave steepening \citep{Hada89,Kennel88:jetp,Malkov91,Medvedev97:pop}, and then are transported by plasma flows having spatially and temporally varying flow speed and Alfven speed. Such variations of plasma flow speed and Alfven speed would force discontinuity-discontinuity interactions, with the possible triggering of the magnetic reconnection and plasma mixing. This scenario is actively discussed for the solar wind turbulence \citep[see, e.g., in][]{Greco08,Greco09,Servidio11,Servidio11:npg,Pecora18,Voeroes21}, and it can be quite perspective for magnetized flows of pulsar wind carring multiple discontinues  \cite[e.g.,][]{Coroniti90,Bogovalov99,Spitkovsky06,Cerutti&Giacinti20A&A}. For pulsar wind the magnetic reconnection is considered as an important mechanism of magnetic energy transformation to plasma energy with the magnetic field reduction and formation of pulsar flares \citep{Arons12,Abdo11,Aharonian12}. However, only internal instabilities of discontinuities and their interaction with the shock waves have been considered as a trigger of the magnetic reconnection \cite[e.g.,][]{Lyubarsky01,Uzdensky11,Cerutti12,Cerutti17}. A proposed here discontinuity-discontinuity interaction may be an alternative mechanism for the  magnetic reconnection triggering, that would work for generally stable rotational discontinuities and would not require magnetic reconnection localization around the shock wave. 

To conclude, we have described the magnetic reconnection triggering due to the interaction of two discontinuities: the solar wind rotational discontinuity (modified and intensified via the bow shock crossing) and tangential discontinuity of the Earth's magnetopause. Spacecraft observations and global hybrid simulations demonstrate that magnetic reconnection triggered by the solar discontinuity impact on the magnetopause results in mixing of hot rarefied magnetospheric plasma and cold dense solar wind (magnetosheath) plasma. We also show that such discontinuity-driven reconnection can be observed at both day-side magnetopause (where solar wind plasma flow contributes significantly to the pressure balance across the magnetopause) and the deep flank magnetopause (where solar wind plasma flow is almost tangential to the magnetopause surface). These results underline an importance of discontinuity-discontinuity interactions for plasma mixing via magnetic field line reconnection.

\section*{Acknowledgements}
\noindent Alexander Lukin and Anatoli Petrukovich acknowledge support from Russian Science Foundation through grant No. 19-12-00313 covering the covering general spacecraft data analysis and data/model comparison parts of this work.
\noindent Anton Artemyev acknowledges support from NASA HSR 80NSSC20K1788 and HGI 80NSSC22K0752 covering analysis of discontinuity configuration aspect of this work.

Anton Artemyev and Xiaojia Zhang also acknowledge NASA contract NAS5-02099 for use of data from the THEMIS mission.  We thank K. H. Glassmeier, U. Auster, and W. Baumjohann for the use of FGM data provided under the lead of the Technical University of Braunschweig and with financial support through the German Ministry for Economy and Technology and the German Aerospace Center (DLR) under contract 50 OC 0302. THEMIS data is available at \url{http://themis.ssl.berkeley.edu}. Data access and processing was done using SPEDAS V4.1 \citep{Angelopoulos19} available at \url{https://spedas.org/}.


\end{document}